\documentclass[superscriptaddress,aps,journal=prl,preprint]{revtex4-2}
\usepackage{graphicx}
\usepackage{epstopdf}
\usepackage{bm,amsmath,amssymb} 
\usepackage{color, soul} 
\usepackage{dcolumn}   
\usepackage{multirow}
\usepackage[colorlinks=true, linkcolor=black, citecolor=black, urlcolor=blue]{hyperref}

\newcommand{\note}[1]{{\color{black}{{#1}}}}
\newcommand{\OP}{O$^{p}$}
\newcommand{\ONP}{O$^{np}$}

\usepackage{lineno}
\begin{document}
\title{Ultrahigh oxygen ion mobility in ferroelectric hafnia}
\author{Liyang Ma}
\thanks{These two authors contributed equally}
\affiliation{Key Laboratory for Quantum Materials of Zhejiang Province, Department of Physics, School of Science, Westlake University, Hangzhou, Zhejiang 310024, China}
\author{Jing Wu}
\thanks{These two authors contributed equally}
\affiliation{Key Laboratory for Quantum Materials of Zhejiang Province, Department of Physics, School of Science, Westlake University, Hangzhou, Zhejiang 310024, China}
\author{Tianyuan Zhu}
\affiliation{Key Laboratory for Quantum Materials of Zhejiang Province, Department of Physics, School of Science, Westlake University, Hangzhou, Zhejiang 310024, China}
\affiliation{Institute of Natural Sciences, Westlake Institute for Advanced Study, Hangzhou, Zhejiang 310024, China}
\author{Yiwei Huang}
\affiliation{School of Engineering, Westlake University, Hangzhou, Zhejiang 310030, China}
\author{Qiyang Lu}
\affiliation{School of Engineering, Westlake University, Hangzhou, Zhejiang 310030, China}
\affiliation{Research Center for Industries of the Future, Westlake University, Hangzhou, Zhejiang 310030, China}
\author{Shi Liu}
\email{liushi@westlake.edu.cn}
\affiliation{Key Laboratory for Quantum Materials of Zhejiang Province, Department of Physics, School of Science, Westlake University, Hangzhou, Zhejiang 310024, China}
\affiliation{Institute of Natural Sciences, Westlake Institute for Advanced Study, Hangzhou, Zhejiang 310024, China}

\date{\today}

\begin{abstract}
Ferroelectrics and ionic conductors are important functional materials, each supporting a plethora of applications in information and energy technology. 
The underlying physics governing their functional properties is ionic motion, and yet studies of ferroelectrics and ionic conductors are often considered separate fields. 
Based on first-principles calculations and deep-learning-assisted large-scale molecular dynamics (MD) simulations, we report ferroelectric-switching-promoted oxygen ion transport in HfO$_2$, a wide-band-gap insulator with both ferroelectricity and ionic conductivity. Applying a unidirectional bias can activate multiple switching pathways in ferroelectric HfO$_2$, leading to polar-antipolar phase cycling that appears to contradict classical electrodynamics. This apparent conflict is resolved by the geometric-quantum-phase nature of electric polarization that carries no definite direction. 
Our MD simulations demonstrate bias-driven successive ferroelectric transitions facilitate ultrahigh oxygen ion mobility at moderate temperatures, highlighting the potential of combining ferroelectricity and ionic conductivity for the development of advanced materials and technologies.    
\end{abstract}

\maketitle
\begin{figure}
	\begin{center}
		\includegraphics[width=0.8\textwidth]{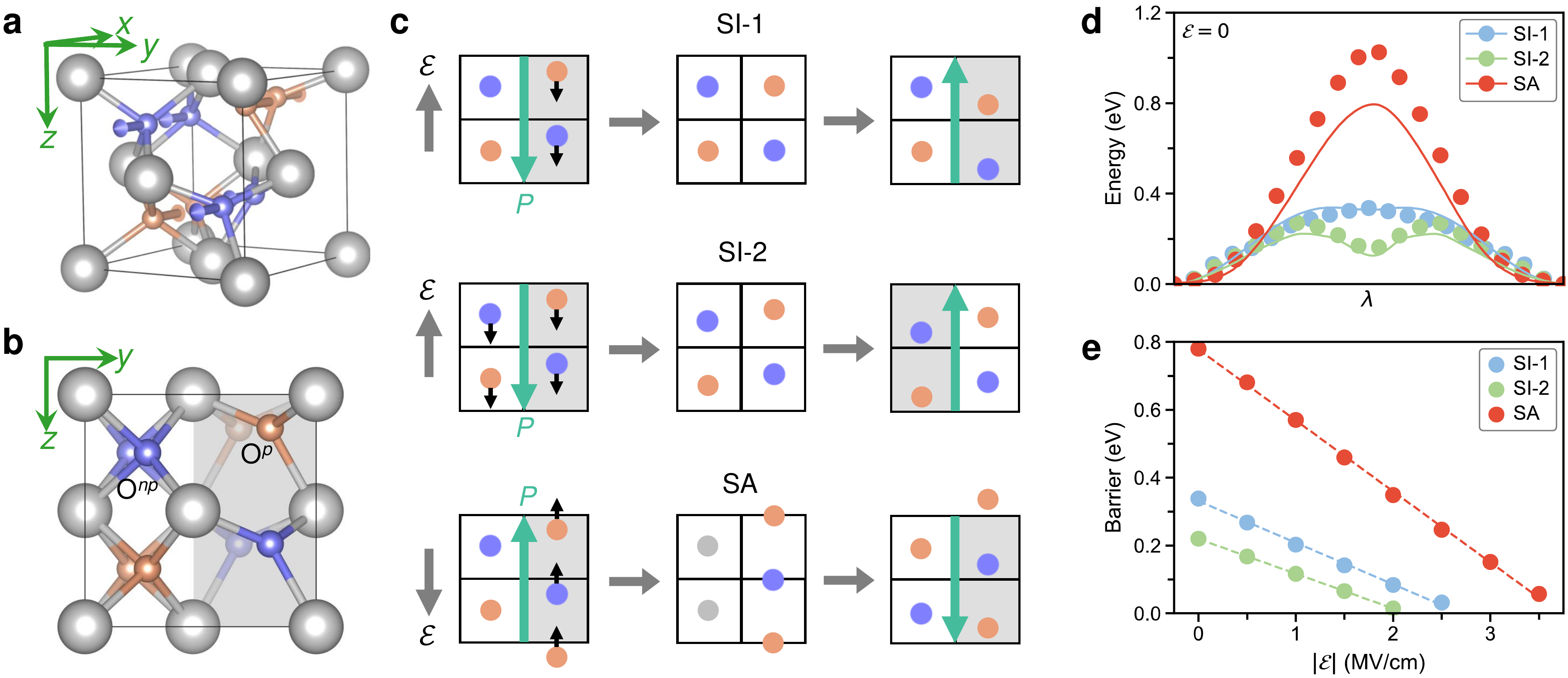}
	\end{center}
	\caption{\textbf{Polarization switching pathways in ferroelectric HfO$_2$.} \textbf{a} X$_2^-$ mode in the unit cell of $Pca2_1$ HfO$_2$ with outward- and inward-displaced oxygen atoms denoted by purple and salmon spheres, respectively. The polarization is along the $z$-axis. \textbf{b} Alternately arranged nonpolar oxygen ions (O$^{np}$) and polar oxygen ions (O$^{p}$) in $Pca2_1$ HfO$_2$. The grey shaded area marks the polar region. \textbf{c} Schematics of shift-inside (SI) and shift-across (SA) switching pathways driven by an external electric field ($\mathcal{E}$). The SA pathway has \ONP~ions reversing the sign of the X$_2^-$ mode (colored in gray during the transition). The SI and SA pathways can start from the same configuration which should be identified by polarization ($P$) vectors (represented as green arrows) pointing in opposite directions to ensure compatibility with classical electrodynamics. \textbf{d} Calculated minimum energy paths for different switching pathways with DFT (lines) and a deep neural network-based force field (scatters). \textbf{e} Switching barrier as a function of field strength.
 \label{pathways}}
\end{figure}

\newpage
Owing to the robust nanoscale ferroelectricity and industry-validated silicon compatibility, HfO$_2$-based ferroelectrics have emerged as an excellent choice for incorporating ferroelectric functionalities into integrated circuits~\cite{Boscke11p102903,Kim21peabe1341}. 
The observed ferroelectricity in hafnia thin films has been attributed to the $Pca2_1$ phase, which is higher in energy than the ground-state monoclinic ($M$) phase. A striking structural characteristic of this polar orthorhombic phase is the presence of a spacing layer consisted of fourfold-coordinated nonpolar oxygen ions (\ONP) that separates polar threefold-coordinated oxygen ions (\OP), and the polar and nonpolar oxygen ions are ordered alternately along the direction perpendicular the polarization ($P$, see Fig.~\ref{pathways}\textbf{a}-\textbf{b}). 

Long before the discovery of ferroelectric HfO$_2$, nonpolar oxygen-deficient hafnia, HfO$_{2-x}$, was actively investigated as a resistive switching material for nonvolatile resistive random access memory~\cite{CervoSulzbach21p3657} where the reversible formation and disruption of conducting filaments composed of chain-like oxygen vacancies are considered to be critical~\cite{Traore15p4029}. Thus, HfO$_2$ is a material system that supports ferroelectricity and ionic conductivity, with both phenomena involving the motion of oxygen ions \cite{Nukala21p630}. Specifically, the polarization switching in $Pca2_1$ HfO$_2$ is characterized by the collective and coordinated local motions of oxygen ions driven by an external electric field ($\mathcal{E}$), whereas the ionic conductivity of HfO$_{2-x}$ features thermally-excited long-distance travel of oxygen ions. Since hafnia films as thin as $\approx$1 nanometer can still retain ferroelectric properties~\cite{Cheema20p478}, and applying voltages of a few volts across such films can generate giant electric fields (up to 9 MV/cm)~\cite{Song20p3221}, exploring the potential interplay between ferroelectric switching and ion transport at high fields is thus important for developing reliable, ultra-dense HfO$_2$-based nanoelectronics, and could also offer insights for the design of field-assisted fast ion conductors. 

The atomistic mechanism of polarization switching in $Pca2_1$ HfO$_2$ remains illusive, partly due to the unusual structural characteristic discussed above and the existence of multiple switching pathways~\cite{Choe21p8,Wei22p154101}. A useful guide is the X$_2^-$-mode-matching criterion. The X$_2^-$ lattice mode features antiparallel $x$-displacements of neighboring oxygen ions perpendicular to the polar axis along $z$ (Fig.~\ref{pathways}\textbf{a-b}), and a pathway conserving the sign of X$_2^-$ mode generally has a lower barrier~\cite{Ma23p096801}. 
 The switching pathways in HfO$_2$ at the unit cell level can be categorized as shift-inside (SI) and shift-across (SA).
As shown in Fig.~\ref{pathways}\textbf{c}, the SI pathways have oxygen ions moving between two Hf atomic planes. Specifically, the SI-1 pathway only involves the displacement of \OP~against $\mathcal{E}$, and the transition state acquires a tetragonal phase (space group $P4_2/nmc$); the SI-2 pathway has both \OP~and \ONP~atoms moving against $\mathcal{E}$, resulting in concerted \OP$\rightarrow$\ONP~and \ONP$\rightarrow$\OP. In comparison, \OP~ions move across Hf planes in the SA pathway, accompanied by the X$_2^-$ mode reversal of \ONP~ions. The switching barriers calculated with the variable-cell nudged elastic band (VCNEB) technique based on density functional theory (DFT) are 0.39, 0.22, and 0.79 eV per unit cell (u.c.) for SI-1, SI-2, and SA, respectively (see computational details in below). These values are reproduced by a deep neural network-based classical force field of HfO$_2$ (Fig.~\ref{pathways}\textbf{d}) that is used for MD simulations in this work \note{(Supplementary Sect.~I)}. The $\mathcal{E}$-dependent switching barriers estimated with VCNEB zero-field barriers are displayed in Fig.~\ref{pathways}\textbf{e}. We find that the critical switching fields (which reduce the barriers to zero) range from 2--4 MV/cm, consistent with experimentally observed coercive fields (1--5 MV/cm)~\cite{Song20p3221,Ku22p154039,Kim21peabe1341}. As we will discuss further, MD simulations employing a large supercell of HfO$_2$ consisting of 28,800 atoms confirm that all three mechanisms are activated at room temperatures when exposed to an electric field of a strength relevant to thin-film device operating conditions. Moreover, applying a unidirectional bias can drive successive ferroelectric switching that supports a continuous flow of oxygen ions even in the absence of oxygen vacancies.

All first-principles DFT calculations are performed using Vienna \textit{ab initio} simulation package  (\texttt{VASP})~\cite{Kresse96p11169, Kresse96p15} with Perdew-Burke-ErnZerhof (PBE) density functional~\cite{Perdew96p3865}. The optimized lattice constants of $Pca2_1$ HfO$_2$ are $a=5.266$ {\AA}, $b=5.048$ {\AA}, and $c=5.077$ {\AA}, and the polarization is along the $c$-axis ($z$-axis). The polarization switching pathways reported in Fig.~\ref{pathways}\textbf{d} are based on a 12-atom unit cell consisting of four hafnium and eight oxygen atoms. The minimum energy paths (MEPs) of SI-1, SI-2 and SA processes are determined using the VCNEB technique implemented in the \texttt{USPEX} code~\cite{Oganov06p244704,Lyakhov13p1172,Oganov11p227}, during which the lattice constants are allowed to relax. The plane-wave cutoff is set to 600 eV. A $4\times4\times4$ Monkhorst-Pack $k$-point grid is used for structural optimizations and VCNEB calculations. 
The stopping criterion for searching the MEP is when the root-mean-square forces on images are less than 0.03 eV/{\AA}. The variable elastic constant scheme is employed in VCNEB, and the spring constant between neighboring images is set within a range of 3.0 to 6.0 eV/\AA$^2$. Energy and polarization values for configurations along the MEP are calculated, with polarization determined using the Berry phase method. The zero-field energy profile for a MEP is subsequently corrected by the $-P\cdot \mathcal{E}$ term,providing an estimated switching barrier under a specific $\mathcal{E}$. To investigate the intrinsic mechanisms of field-driven ferroelectric switching in $Pca2_1$ HfO$_2$, a defect-free single-domain supercell with 12,000 atoms is chosen as the initial configuration for MD simulations. We perform isobaric-isothermal ensemble ($NPT$) MD simulations over a wide range of electric fields from 0 to 12 MV/cm at 400 K, 500 K and 600 K, utilizing a deep neural network-based force field. The model potential is obtained by deep learning from a database of energies and atomic forces for $\approx$55,000 configurations computed with DFT (see details in Supplementary Sect.~I)~\cite{Wu23p144102}. All $NPT$ MD simulations are carried out using LAMMPS~\cite{Plimpton95p1}, with temperature controlled via the Nos\'{e}-Hoover thermostat and the pressure controlled by the Parrinello-Rahman barostat. The integration timestep for the equation of motion is 1 fs in all MD simulations. At a given temperature, the equilibrium run is 20 ps with pressure maintained at 1.0 bar, followed by a production run of 500 ps at the specified temperature and electric field, ensuring reliable estimation of mean square displacement (MSD) of all oxygen ions and the mobility $u_{\rm O}$ (Supplementary Sect.~IV).

\begin{figure}
	\begin{center}
		\includegraphics[width=0.8\textwidth]{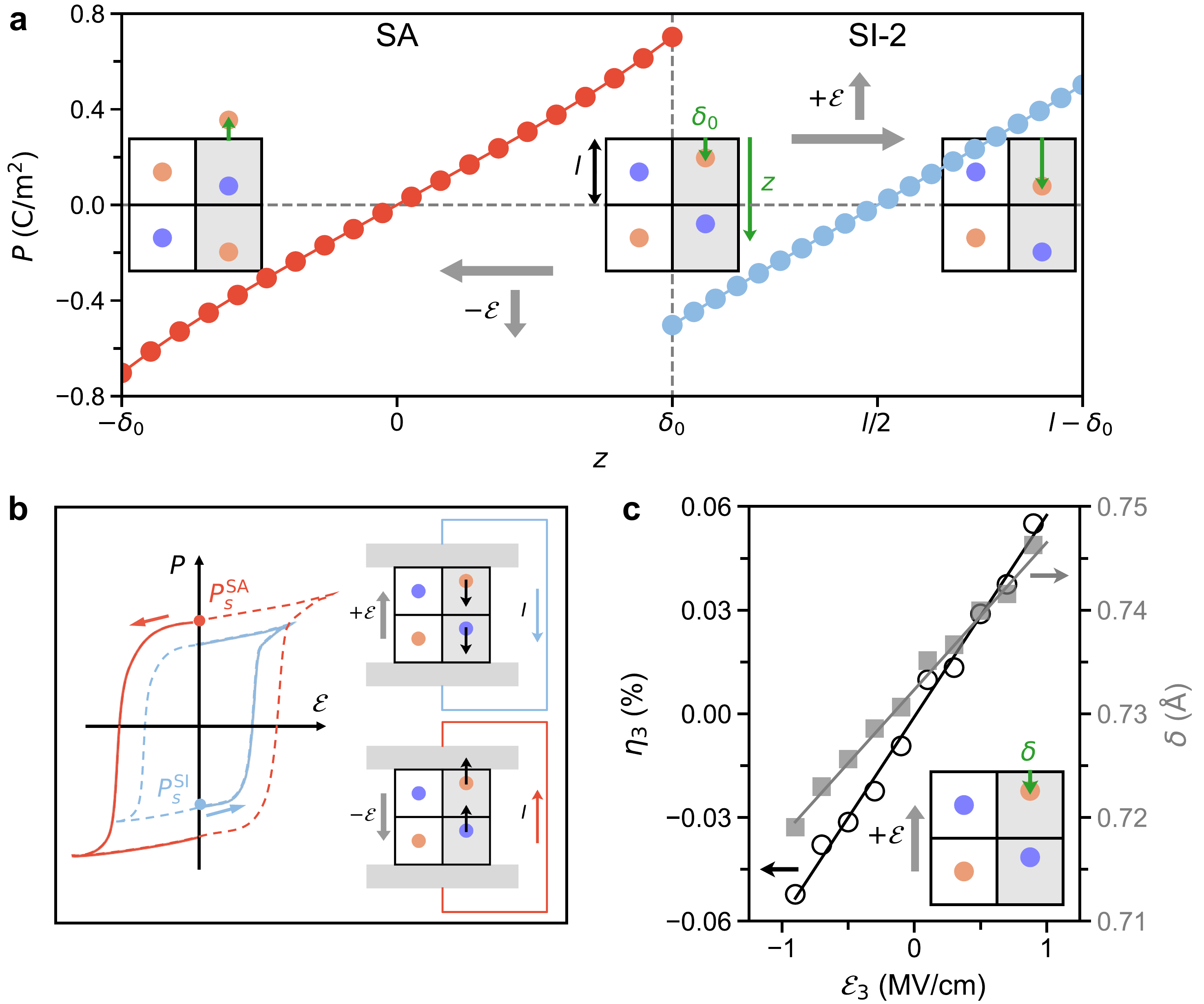}
	\end{center}
	\caption{\textbf{Dual-valued remnant polarization and single-valued piezoelectric response in HfO$_2$.} \textbf{a} Polarization variation along SA and SI-2 switching pathways from the same starting configuration (the center insert) in response to opposing electric fields. The upward electric field aligned along the $-z$ direction is defined arbitrarily as $+\mathcal{E}$. Configurations are labeled by the displacement ($\delta$) of the salmon-colored \OP~ion relative to the top Hf plane; $l$ is the distance between neighboring Hf planes along $z$ and $\delta_0$ is the \OP~displacement at the ground state. Oxygen ions always move against $\mathcal{E}$.
 \textbf{b} Schematics of $P$-$\mathcal{E}$ hysteresis loops for SA and SI and the corresponding switching currents. \textbf{c} Strain ($\eta_3$, empty circles) as a function of an electric field applied along the $z$-axis ($\mathcal{E}_3$) and the corresponding \OP~displacements ($\delta$, filled squares)~obtained with finite-field MD simulations at 300 K. The slop of $\eta_3$-$\mathcal{E}_3$ gives the absolute value of $d_{33}$ whose sign depends solely on the arbitrary sign of $\mathcal{E}_3$.
 \label{polarization}}
\end{figure}

Upon closely examining the SI and SA pathways, a perplexing behavior becomes evident. For the same starting configuration depicted in Fig.~\ref{pathways}\textbf{c}, external electric fields in opposing directions can both drive ferroelectric switching. Consequently, in order to conform with classical electrodynamics, the same configuration would exhibit a downward polarization in SI but an upward polarization in SA. 
We emphasize that the macroscopic electric polarization of a crystalline solid is a geometric quantum phase, which should be viewed as a multi-valued lattice property with no definite direction ~\cite{Smith93p1651,Vanderbilt93p4442}. However, for practicality and compatibility with classical electrodynamics, electric polarization is often treated as a vector with a specific direction. We calculate the polarization with the Berry phase approach by tracking the Berry phase variation during SI-1 and SA pathways. The results for SI-2 are similar to SI-1 \note{(Supplementary Fig.~S2)}. Here, the upward electric field is defined  (arbitrarily) as $+\mathcal{E}$ that aligns along the $-z$ direction. As illustrated in Fig.~\ref{polarization}\textbf{a}, SI-1 and SA pathways correspond to two different branches of the polarization lattice, each associated with a definite change in polarization ($\Delta P$) without ambiguity. The magnitude of $\Delta P_{\rm SI}$ for the SI-1 pathway driven by $+\mathcal{E}$ is 1.0 C/m$^2$, and the polar state of the initial configuration can be {\em labeled} as $P_s^{\rm SI}=-0.5$~C/m$^2$ to be consistent with classical electrodynamics. Similarly, the SA pathway driven by $-\mathcal{E}$ results in $|\Delta P|$ of 1.4~C/m$^2$, and we can label the same starting configuration as $P_s^{\rm SA}=+0.7$~C/m$^2$. Because the polarization change in each pathway is well defined and can be connected to experimentally measurable observables such as switching current, HfO$_2$ is a unique ferroelectric with dual-valued remnant polarization ($P_s^{\rm SI}$ and $P_s^{\rm SA}$) characterized by two intrinsic $P$-$\mathcal{E}$ hysteresis loops (Fig.~\ref{polarization}\textbf{b}). We note that experimentally giant polarization magnitudes of 0.5--0.64~C/m$^2$ in polycrystalline films of hafnia have been reported ~\cite{Ku22p154039,Yun22p903}, suggesting the realization of SA switching mechanism and $P_s^{\rm SA}$.

Because all ferroelectrics are piezoelectric and piezoelectricity is typically gauged by the piezoelectric strain coefficient ($d$) that links strain ($\eta$) and $\mathcal{E}$ via $\eta_i=d_{ij}\mathcal{E}_j$, an interesting question arises: does ferroelectric HfO$_2$ with two remnant polarization values (switching pathways) also possess two values of $d_{33}$? We discover that despite the dual-valued nature of $P_s$, HfO$_2$ exhibits an unambiguous piezoelectric response, as hinted by the parallel SA and SI branches with identical slopes in Fig.~\ref{polarization}\textbf{a}.
Our finite-field MD simulations reveal that an electric field applied along the $z$-axis ($\mathcal{E}_3$) that drives~\OP~ions away from the nearest Hf atomic plane leads to lattice expansion ($\eta_3>0$) and vice versa (Fig.~\ref{polarization}\textbf{c}).
Consequently, for a given crystal orientation and ${\mathcal{E}_3}$, the field-induced $\eta_3$ is unique; the absolute value of $d_{33}$ computed with $|\partial{\mathcal{E}_3}/\partial{\eta_3}|$ is single-valued, while the sign of $d_{33}$ depends solely on the sign of $\mathcal{E}_3$ (the arbitrary choice of the positive field direction). The estimated $|d_{33}|$ is 5.83~pm/V, comparable with both DFT ($2.59$ pm/V) ~\cite{Liu20p197601} and experimental (2--5 pm/V)~\cite{Dutta21p7301} values. 

\begin{figure}
	\begin{center}
		\includegraphics[width=0.8\textwidth]{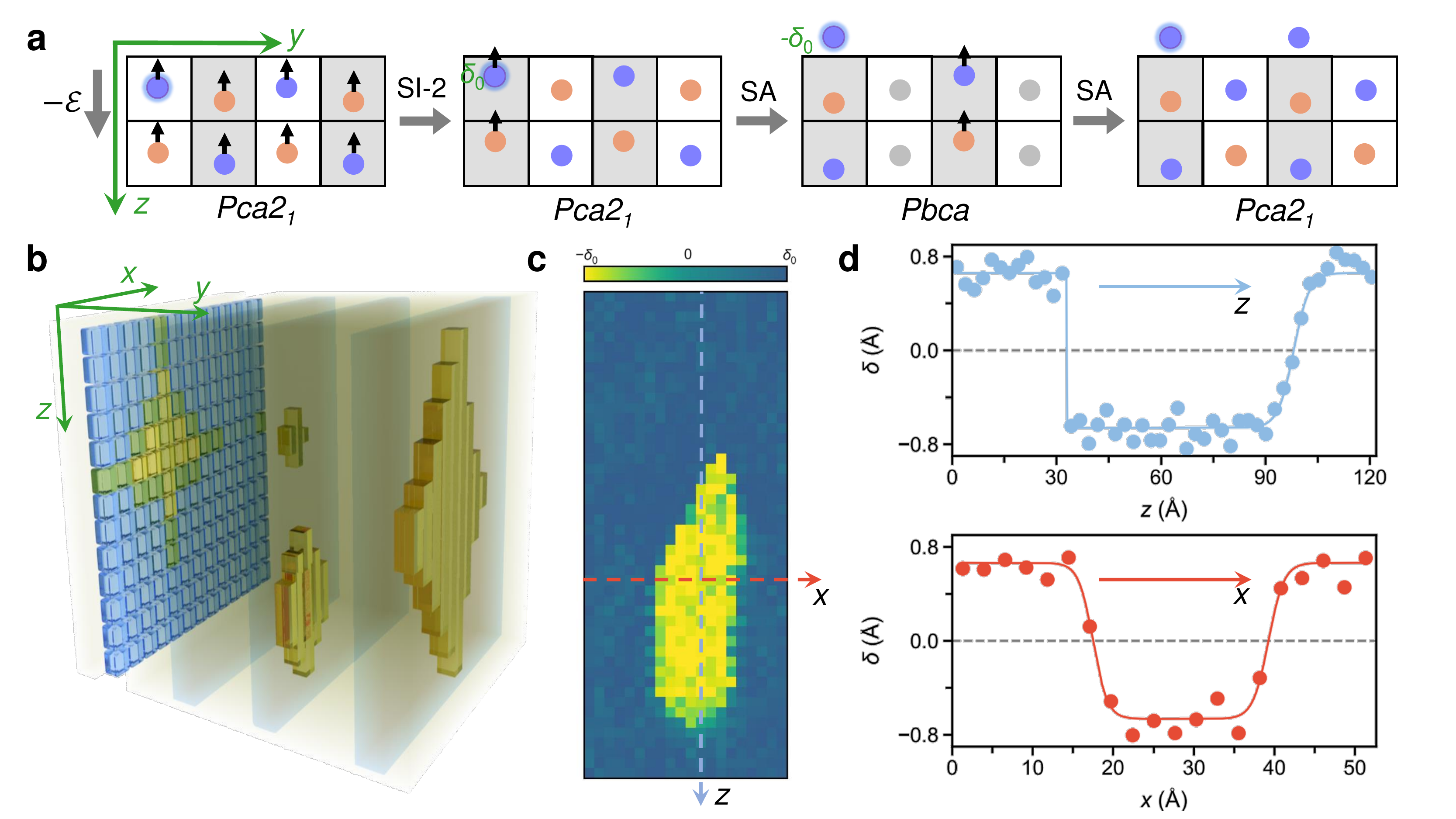}
	\end{center}
	\caption{\textbf{Oxygen ion transport coupled to the nucleation-and-growth mechanism of ferroelectric switching.} \textbf{a} Polar-antipolar phase cycling arising from successive SI and SA ferroelectric transitions. The highlight \ONP~in the initial configuration becomes \OP~with $\delta=\delta_0$ after SI-2 and then \OP~with $\delta=-\delta_0$ after SA. \textbf{b} Schematic illustration of stochastic nucleation events  in a three-dimensional (3D) bulk. The nucleus is two dimensional (2D) within the $xz$-plane, featuring a thickness equivalent to half a unit cell of $Pca2_1$ HfO$_2$ along the $y$-axis. \textbf{c} A 2D slim-diamond-shaped nucleus extracted from MD simulations using a $10\times10\times24$ supercell of 28,800 atoms. The nucleus profile is determined based on the $\delta$ values of \OP~ions. \textbf{d} Line profiles of $\delta$ along the $z$ and $x$ directions marked in \textbf{c}.
	\label{MD}}
\end{figure}

Importantly, the process of oxygen ions traversing unit cells by SA following SI can be viewed as a classical analogue of adiabatic Thouless pumping, and can be achieved by applying a constant bias. We perform large-scale finite-field MD simulations and confirm that a unidirectional $\mathcal{E}$ can indeed drive successive SI and SA ferroelectric transitions that support a continuous flow of oxygen ions even in the absence of oxygen vacancies. 
Figure~\ref{MD}\textbf{a} illustrates a typical local switching process extracted from MD simulations. The initial configuration has \OP~ions situated near the bottom Hf planes, and a negative $\mathcal{E}$ (aligned along $+z$) drives the SI pathway during which negatively charged oxygen ions move against $-\mathcal{E}$. 
Notably, unit cells can further transform to an antipolar $Pbca$ phase and subsequently undergo another transition from $Pbca$ back to $Pca2_1$ under the same bias, each through the SA mechanism. Locally, unit cells return to their original configuration albeit translated by half of a $Pca2_1$ unit cell along the $y$-axis. 
This phase cycling would be difficult to comprehend if a fixed polarization direction is assigned to a particular crystal configuration; it is again a manifestation of the geometric-quantum-phase nature of electric polarization that does not possess a definite direction. Microscopically, this phenomenon is a natural consequence of continuous flow of oxygen ions against the direction of the applied external electric field.

We find that the transport of oxygen ions is directly coupled to the nucleation-and-growth mechanism of ferroelectric switching. 
The SA step that is associated with a larger barrier than SI (Fig.~\ref{pathways}\textbf{c}) serves as the rate limiting step. The nucleus is then characterized by a domain of unit cells that have completed the SA step. 
MD simulations reveal several microscopic features of the nucleation-and-growth mechanism, sketched in Fig.~\ref{MD}\textbf{b}. 
First, 
even though the switching process occurs in a three-dimensional (3D) bulk, the nucleus formed in the presence of $-\mathcal{E}$ is nearly two dimensional (2D), as opposed to small 3D clusters typically observed in ferroelectric perovskites \cite{Prosandeev18p024105}. The nucleus has a thickness of merely half a unit cell along the $y$-axis and assumes a slim diamond shape in the $xz$ plane (Fig.~\ref{MD}\textbf{c}). 
This is surprisingly similar to the nucleus formed at a moving domain wall in perovskite ferroelectrics~\cite{Liu16p360}. The ability to form a 2D nucleus in 3D can be attributed to the weak dipole-dipole interactions along the $y$-axis resulting from \ONP~spacing layers ~\cite{Lee20p1343}. Second, the nucleus exhibits anisotropic diffusive interfaces.
The nucleus profile is determined based on the displacement ($\delta$) of \OP~ions, and the unit cells with SA completed have $\delta=-\delta_0$ (Fig.~\ref{MD}\textbf{a}). The interfacial profile is fitted to $\delta_0 \tanh\left(\frac{\delta}{\gamma_i/2}\right)$ with $\gamma_i$ characterizing the diffusiveness of the
nucleus along direction $i$ ($i$=$x$,$z$).
As presented in Fig.~\ref{MD}\textbf{d}, the longitudinal diffusiveness parameter, $\gamma_z$, is 8.9~\AA~at one side but becomes zero at the other side. In comparison, the lateral diffusiveness parameters, $\gamma_x$, are roughly of the same value (3.7~\AA) at both sides. The considerable diffusiveness reduces the interface energy, which in turn decreases the nucleation barrier. 
Lastly, nucleation events exhibit stochastic behavior. Nuclei of varying sizes randomly emerge throughout the system, and only those exceeding a critical size continue to expand, eventually leading to the switch of the entire $xz$ layer (Supplementary Fig.~S4). 

\begin{figure}
	\begin{center}
		\includegraphics[width=0.6\textwidth]{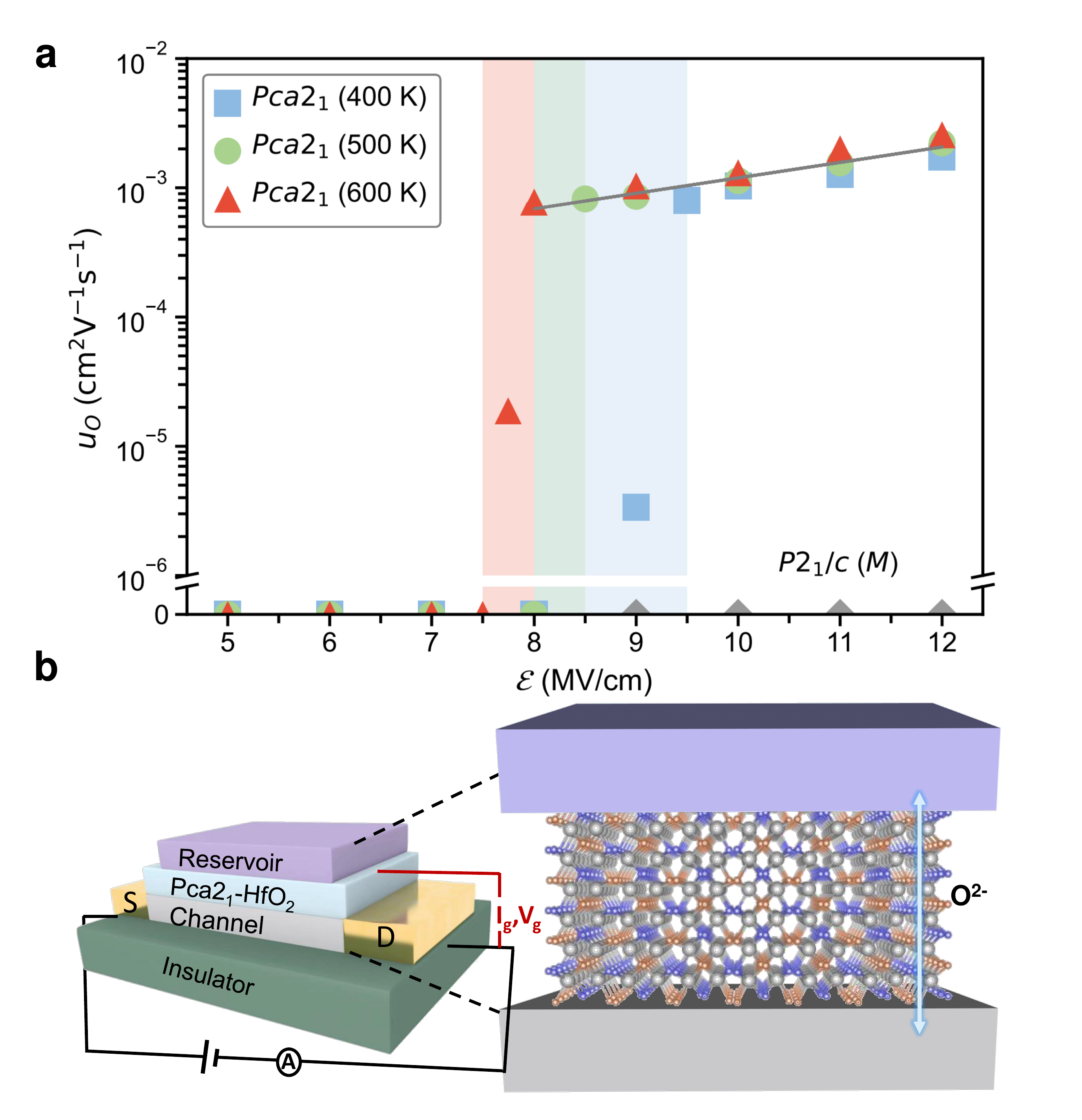}
	\end{center}
	\caption{\textbf{Ferroelectricity-promoted oxygen ion mobility in HfO$_2$.} \textbf{a} Mobility of oxygen ions ($u_{\rm O}$) in $Pca2_1$ HfO$_2$ as a function of $\mathcal{E}$ at different temperatures from MD simulations. The results in nonpolar $M$ phase at 600 K are shown for comparison. The shaded area indicates the transition region where the critical electric field $\mathcal{E}_t$ is located. \textbf{b} Left: Device structure of an electrochemical ionic synapse with $Pca2_1$ HfO$_2$ as the electrolyte layer, which is sandwiched between a channel layer connected to source (S) and drain (D) electrodes and an oxygen-storing reservoir layer. The channel layer has its conductance depending on the oxygen ion concentration. Right: Schematic showing the ultrafast oxygen ion transport in HfO$_2$ electrolyte layer under an electric field.
	\label{cycle}}
\end{figure}

We quantitatively estimate the mobility of oxygen ions in defect-free HfO$_2$ under moderate temperatures and over a range of electric fields using MD simulations. By utilizing a vacancy-free model, the occurrence of vacancy-mediated ion diffusion processes is eliminated. Figure~\ref{cycle}\textbf{a} plots the mobility of oxygen ion ($u_{\rm O}$) in $Pca2_1$ as a function of $\mathcal{E}$ at 400, 500, and 600~K, respectively, compared to that in the nonpolar $M$ phase at 600~K.
The $u_{\rm O}$-$\mathcal{E}$ relationships in $Pca2_1$ reveal a temperature-dependent critical field ($\mathcal{E}_{t}$), below which the mobility is strictly zero because only local SI switching events are activated. 
Above $\mathcal{E}_{t}$, $u_{\rm O}$ quickly jumps to a giant value of $\approx$$10^{-3}$~cm$^{2}$V$^{-1}$s$^{-1}$. We observe that an increase in temperature leads to a reduction in the magnitude of $\mathcal{E}_{t}$ which represents the field required to activate the SA step. Interestingly, $u_{\rm O}$ shows a weak temperature dependence when above $\mathcal{E}_{t}$ and mainly depends on the strength of the driving field, indicating a depinning-like behavior \cite{Liu16p360, Tuckmantel21p117601}. 
In comparison, the value of $u_{\rm O}$ in $M$ at 600 K remains strictly zero due to the absence of ferroelectricity. These results demonstrate 
 the considerable influence of ferroelectricity on oxygen ion mobility in HfO$_2$, particularly in the high-field region.

One of the potential applications of hafnia thin films with ultrahigh oxygen ion mobility enabled by successive ferroelectric switching is electrochemical ionic synapses (EIS) based on oxide ion migration, which are emerging neuromorphic computing devices for artificial neural networks. EIS devices function like nano-batteries, utilizing ion migration for computing-in-memory operations. An EIS device structure is shown in Fig.~\ref{cycle}\textbf{b}, which includes a channel layer with conductance that varies based on oxygen ion concentration, an oxygen-storing reservoir, and an electrolyte layer connecting the channel and reservoir for oxygen ion migration~\cite{Huang22p2205169,Onen22p539}. The conductance of the channel material can be modulated step-wise by applying an electrical bias across the tri-layer device which triggers the oxygen ion transport. 
Therefore, the ultrahigh oxygen ion mobility in silicon-compatible ferroelectric HfO$_2$ under an electric field can potentially enable scalable EIS devices with ultrafast speed.

In summary, this study highlights the geometric-quantum-phase attribute of spontaneous electric polarization in ferroelectric $Pca2_1$ HfO$_2$ that displays dual-valued remnant polarization and single-valued piezoelectric response.
Successive ferroelectric switching, driven by constant bias and resembling Thouless pumping, can boost oxygen ion transport at moderate temperatures. Microscopically, the long-distance travel of oxygen ions is directly coupled to the nucleation-and-growth mechanism. Similar phenomena may occur in other ferroelectric systems that support successive switching pathways such as CuInP$_2$S$_2$ and LaVO$_3$-SrVO$_3$ superlattice~\cite{Qi22p06999}. The integration of ferroelectricity and ionic conductivity unlocks new possibilities for innovative device types, including ferro-electrochemical ionic synapses.

\begin{acknowledgments}
L.M., J.W., T.Z., and S.L. acknowledge the supports from National Key R\&D Program of China (2021YFA1202100), National Natural Science Foundation of China (12074319), and Westlake Education Foundation. Y.H. and Q.L. acknowledge funding support from the Research Center for Industries of the Future at Westlake University and National Natural Science Foundation of China (NSFC, Grant No. 52202148). The computational resource is provided by Westlake HPC Center. 
\end{acknowledgments}

\newpage

\newpage

\newpage

\newpage

\clearpage

\bibliography{SL.bib}

\end{document}